\documentclass{article}%
\usepackage{lineno}
\usepackage{xcolor,colortbl}
\usepackage{eurosym}
\usepackage[top=3cm, bottom=3cm, left=3cm, right=3cm]{geometry}
\usepackage{amsmath}
\usepackage{amsfonts}
\usepackage{amssymb}
\usepackage{graphicx}%
\usepackage{epsfig}
\usepackage{xcolor,colortbl}
\usepackage{float}
\date{}
\begin{document}

\title{\textbf{Quark Model Study of Doubly Heavy $\Xi$ and $\Omega$ Baryons via Deep Neural Network and Hybrid Optimization}}
\author{Zahra Ghalenovi$^{1}$\footnote{$z_{-}ghalenovi@kub.ac.ir$}~, Masoumeh Moazzen Sorkhi$^{1}$ and Amir Hossein Sovizi$^{2}$\\ $^{1}$Department of Physics, Kosar University of Bojnord, Bojnourd 94156-15458, Iran\\ $^{2}$Yadegar Emam, Shahre  Rey Branch, Islamic  Azad University, Tehran 18155-144, Iran}

\maketitle

\begin{abstract}
In the present work we investigate the mass spectrum and semileptonic decays of double charm and bottom baryon states using the hypercentral quark model. We solve the six-dimensional Schrödinger equation via deep learning and particle swarm optimization techniques to improve the speed and accuracy. Then, we predict the masses of the ground and excited states of single and doubly heavy baryons. Working close to the zero recoil point, we also study the semileptonic decay widths and branching ratios of doubly heavy $\Xi$ and $\Omega$ baryons for the $b\rightarrow c$ transitions. A comparison between our results and the evaluations of other theoretical models is also presented. Our predictions of mass spectrum and decay widths provide valuable information for the experiment searching for undiscovered heavy baryon states. 
\end{abstract}

\section{Introduction}

During the last few years various theoretical models have been presented on heavy baryon spectroscopy and decays and study on this topic has been of special interest~\cite{Weng2018,Shaha2017,Garcilazo2016}. 
Several studies such as non-relativistic quark model \cite{Kumar2023, Woloshyn, Ghalenovi:2011zz, Ghalenovi:2013}, relativistic quark model \cite{Capstick1, Ebert5}, Lattice QCD \cite{Mathur}, QCD sum rule \cite{Aliev1, Mao, Liu3,Azizi1, Aliev6, Aliev7}, Quark-diquark model \cite{Ebert6, Thakkar6, Thakkar7} and Faddeev approach \cite{Gerasyuta}  have been done on heavy baryons properties.
In 2002 the SELEX Collaboration reported the first observation of $M_{\Xi ^{++}_{cc}}=3518\pm1.7$ MeV ~\cite{Mattson2002} for the lightest doubly heavy $\Xi ^{++}_{cc}$ baryon state in the decay mode $ \Xi ^{+}_{cc}\to \Lambda_c^+ K^- \pi^+$ ~\cite{Mattson2002}. They confirmed this state in another decay mode $ \Xi ^{++}_{cc} \to p D^+ K^-$~\cite{Ocherashvili2005}. However, FOCUS, BaBar, and Belle collaborations reported negative results in observation of the $ \Xi ^{+}_{cc}$ ~\cite{nfocus, nbabar, nbelle}.
In 2017, the $ \Xi ^{++}_{cc}(3621)$ heavy baryon state was announced through the decay channel $ \Xi ^{++}_{cc} \to  \Lambda_c^+ K^-\pi^+\pi^+ $ with a measured mass $ (3621.40\pm0.72\pm0.14\pm0.27)$ MeV  by the LHCb collaboration ~\cite{LHCb2017}, and later confirmed by the same collaboration in another decay mode $ \Xi ^{++}_{cc} \to   \Xi ^{+}_{c} \pi^+$~\cite{LHCb2018} with a mass of 100 MeV greater than the first measured mass reported by the SELEX collaboration~\cite{Mattson2002}. This subject became one of the hot topics both experimentally and theoretically.  The lifetime of the $ \Xi_c^{++}$ baryon state was measured to be $ \tau( \Xi_{cc}^{++})=(0.256^{+0.024}_{-0.022}\pm 0.014)$ ps~\cite{LHCb20188}. In 2019 and 2021 new searching was done for the mass measurement of $\Xi ^{++}_{cc}$ baryon state on the decay channels  $ \Xi ^{++}_{cc}\to \Lambda_c^+ K^- \pi^+$  and  $ \Xi ^{++}_{cc}\to \Xi_c^+ \pi^- \pi^+$ respectively\cite{LHCb2019, LHCb20212}. Their results confirmed the mass of the $ \Xi ^{++}_{cc}$ baryon state to be around 3620 MeV. The mass of other doubly heavy baryons, the beauty-charm  and double-beauty baryons, have not been measured yet and are waiting to be found in the future.
Recently, the LHCb collaboration has started new searches on the $\Xi_{bc}$ and $\Omega_{bc}$ doubly heavy baryons, but these baryon states are not yet to be observed~\cite{lhcb1,lhcb2}. In comparison to the doubly charmed $\Xi_{cc} $ baryon, searching for the other doubly heavy states are more complicated. The search for these states is really hard for the experiment since higher energy and higher beam luminosity are required to produce them. For the semileptonic decay widths of doubly heavy baryons there are also a limited number of calculations reported by theoretical studies~\cite{Li2023,Shi2020,Gutsche2019,Yu2019,Hernandez2008,Lyubovitskij2003,Albertus2005,Wang2017,Qin2021} and proposing a theoretical model on this topic is of great interest.  The theoretical studies on semileptonic decays of doubly heavy baryons started from 2017 \cite{Li2017, Yu2018} in which the discovery channel of  $ \Xi ^{++}_{cc} \to  \Lambda_c^+ K^- \pi^+ \pi^-$  and  $\Xi ^{+}_{c}  \pi^+ $ is pointed out. The exclusive weak decays of doubly heavy baryons was also studied in Ref. \cite{Gerasimov2019} using the factorization theorem.  The authors of Ref. \cite{Albertus2008} studied the static properties and semileptonic decays of the doubly heavy baryons by using the variational method. In Ref. \cite{Li2020} the wave functions of doubly heavy baryons were obtained by solving Bethe–Salpeter equations and the baryon mass spectrum was calculated. In Refs. \cite{Kumar2023,Ghalenovi:2013} the mass spectra, magnetic moments and weak decays of double heavy baryons were calculated in the hypercentral quark model.\\
In recent years, machine learning has been the most popular computational approach for different fields of modern science and a significant raise of interest has been seen in that area. The machine learning techniques enable us to improve the performance and accuracy in theory and experiment. \\ 
The solution of three-body Schrödinger equation is a complicated problem in hadron physics. The machine learning techniques such as deep neural network (DNN) help us to solve this problem in a short time. Deep learning algorithms can identify the energy eigenvalues of the baryon systems~\cite{Yadav:2015,Parisi:2003,Mutuk: EPJA2020}. 
Use of neural networks for solving differential equations has some advantages \cite{Yadav:2015,Parisi:2003}. One point is that by increasing the dimensions of problem, the computational complexity does not increase considerably. The other advantage is that the solution of the problem is continuous over the domain of integration \cite{Mutuk: EPJA2020}. In this study, we present a mathematical model using the DNN combining with particle swarm optimization algorithm (PSO) which is fast and provides high accuracy in mass estimation of the single and doubly heavy baryons. Some earlier works studied the heavy hadron properties via neural networks and deep learning techniques \cite{Mutuk: EPJA2020,Gal2022,Malekhosseini:PRD2024,Mutuk:CPC2019}.  Ref. \cite{Mutuk: EPJA2020}  solved the nonrelativistic Schrödinger  equation via neural networks and calculated the masses of the heavy $\Omega_b$  baryon states in a hypercentral approach. Ref. \cite{Gal2022} used the neural networks and Gaussian processes to calculate the mass spectra of the light baryons, heavy charm mesons and pentaquark states.  
The authors of Ref. \cite{Malekhosseini:PRD2024} presented two approaches using the deep neural networks to study the masses and decay widths of the mesons. Ref. \cite{Mutuk:CPC2019} solved five-body Schrödinger equation of pentaquark states by using the neural networks and predicted the parities of these states. \\
In the present work, the single and doubly heavy baryon states are studied in a quark model. Firstly, we introduce our model and simplify the six-dimensional Schrödinger equation using the hypercentral model. Then, by applying the deep learning techniques, we evaluate the energy eigenvalues of the considered baryon system.  In our method, we also use the particle swarm optimization to get the best values of the energies. In the next step, we predict the mass spectra of the ground and excited states of the single and doubly heavy baryons and present a comparison between our results and the experimental measurements or other theoretical predictions. The exist experimental data of the single heavy baryon masses help us to optimize our mass evaluations of the $ \Xi $ and $ \Omega $ doubly heavy baryons. Finally, we simplify the form factors of $ b\rightarrow c $ semileptonic decays and evaluate the semileptonic decay widths and branching ratios of the doubly heavy $ \Xi $ and $ \Omega $ baryons.  It helps us to get a deep understanding of their structure. \\
The structure of the present work is as follows. In Section~\ref{model} we introduce the three-body hypercentral quark model and present our potential model. In Section~\ref{NN model} the deep neural networks is briefly introduced. Our calculations of the single and doubly heavy baryon masses are presented in Section~\ref{mass}.  In Section~\ref{semi decay} we simplify the form factors of the considered transitions and calculate the semileptonic decay widths and corresponding branching fractions of doubly charm and bottom heavy baryons. A short summary is given in Section~\ref{Summary}.

\section{Phenomenological model} \label{model}
In the heavy-quark sector of heavy baryons one can use a non-relativistic formalism to study these states. The configurations of three quarks of baryons can be described by the Jacoobi coordinates, $ \rho $ and $ \lambda $,
\begin{equation}  \label{rhoo}%
\vec{\rho} = \frac{1}{\sqrt{2}}(\vec{r_1} - \vec{r_2}), \qquad
\vec{\lambda} = \frac{1}{\sqrt{6}}(\vec{r_1} + \vec{r_2} - 2
\vec{r_3})  
\end{equation}
such that
\begin{equation}
m_{\rho} = \frac{2 m_1 m_2}{m_1 + m_2}, \qquad m_{\lambda} =
\frac{2 m_3 (m_1^2 + m_2^2+m_1m_2)}{(m_1+m_2) (m_1+ m_2 + m_3)}    
\end{equation}
Here $m_1$, $m_2$ and $m_3$ are the constituent quark masses.
In the hypercentral model, the hyperradius $x$ and hyperangle $\zeta$ are given in terms of the Jacoobi coordinates  $\rho$ and $\lambda$  \cite{Giannini:1998,Giannini:2}  
by 
\begin{equation}
x = \sqrt{\rho^2 + \lambda^2}, \qquad \xi = \arctan
(\sqrt{\frac{\rho}{\lambda}}).
\end{equation}

The Hamiltonian of the baron system takes the form
\begin{equation}
H=\frac{p^{2}_{\rho}}{2m_{\rho}}+\frac{p^{2}_{\lambda}}{2m_{\lambda}}+V(x).
\end{equation}

The hyperradial baryon wave function $ \psi_{\nu,\gamma}(x) $ is defined as a solution of the following equation (for details see Refs. \cite{Giannini:2,Ghalenovi:2014})
\begin{equation}\label{Schro}
[\frac{d^2}{dx^2}+\frac{5}{x}\frac{d}{dx}-\frac{\gamma(\gamma+4)}{x^2}]\psi_{\nu,\gamma}(x)=-2m[E_{\nu,\gamma}
-V(x)]\psi_{\nu,\gamma}(x).
\end{equation}
where $ m $ is the reduced mass and $\gamma$ is the grand angular quantum number. $\nu$ counts the nodes of the wave function. We consider the Killingbeck potential given by   \cite{Ghalenovi:2014}

 \begin{equation} \label{v(x)}
 V(x)=\alpha x^2+\beta x-\frac{\tau}{x},
 \end{equation}
 
here, $x$ is the hyperradius  and $\alpha $, $\beta$ and  $\tau$ are constant. In the present work, we solve the Schrödinger equation \ref{Schro} using deep learning technique combined with a PSO algorithm. We use a trial wave function as proposed in our previous works~\cite{Ghalenovi:2022dok,Ghalenovi:2018fxh}. In the present study we consider the baryon states with zero nodes $\nu=0 $, and neglect the spin dependent interactions.

\section{Deep Learning Model for Energy Evaluation} \label{NN model} 
 The first step in our framework involves the generation of a synthetic dataset by solving the Schrödinger equation and using the shooting method. The Schrödinger equation for the system under study is expressed as

\begin{equation} \label{Schro2}
\left[ \frac{d^2}{dx^2} + \frac{5}{x} \frac{d}{dx} - \frac{\gamma(\gamma + 4)}{x^2} - 2m\alpha x^2 - 2m\beta x + \frac{2m\tau}{x} \right] \psi_\gamma(x) = -2m E_\gamma \psi_\gamma(x).
\end{equation}

The goal is to compute $E_{\gamma}$ energy eigenvalues such that the boundary conditions for $ \psi_\gamma(x) $ are satisfied. In this step deep learning model is used to predict energy eigenvalues based on the features in dataset.

\subsection{Model Architecture}

 The deep neural networks consist of three layers: an input layer, hidden layers, and output layer. In the present work, we employ a feed forward neural network with one input layer, two hidden layers and one output layer to solve the differential equation \ref{Schro2}. The output layer consists of a single neuron to predict the  $ E_\gamma $ energy eigenvalues.

\subsection{Training Procedure}
The neural network is trained using the mean squared error (MSE) loss function:
    \begin{equation}
    MSE = \frac{1}{n} \sum_{i=1}^{n} (Y_{\text{true}} - Y_{\text{pred}})^2,
    \end{equation}

where $  n$, $ Y_{\text{true}} $ and $ Y_{\text{pred}} $ are the number of data points, actual value for $ i $-th data point and predicted value for the $ i $-th data point.    
We employ the adaptive moment estimation optimizer for adaptive learning rate adjustments during training. The model is trained over a large number of epochs to ensure convergence.
  The dataset is split into 80\% training and 20\% validation subsets.  The root mean squared error (RMSE) and MSE show the validity of our model and its accuracy.

\section{Particle Swarm Optimization (PSO)}
For further refines of energy evaluations, we employ the particle swarm optimization. PSO is an effective method in finding global optima in multi-dimensional search spaces. A swarm of particles 
is initialized randomly, where each particle represents a solution (energy value). The position of each particle is updated iteratively, 
and the fitness of each particle is evaluated using the trained deep learning model. 
The fitness function for each particle is based on the difference between the predicted energy value and the target energy value. 
The goal is to minimize this difference. The global best solution found by PSO algorithm represents the optimal energy estimations.

\section{Heavy baryon mass spectra} \label{mass} 
By using the obtained $E_{\gamma}$ energies, one can evaluate the mass spectra of the baryon systems. The quark masses are taken from our previous work~\cite{Ghalenovi:2022dok}. We consider $m_{q}$=320 MeV, $m_{s}$=440 MeV, $m_{c}$=1600 MeV and $m_{b}$=4670 MeV. 

\begin{table} [h]
\caption{Masses of the ground and excited states of single charm baryons (in GeV). Last column shows the percentage of relative errors between our results and experimental data.}
\label{tab:mass1}
 \begin{center}
{\begin{tabular}{lllllll} \hline \hline  
State& Baryon&Our results& Exp \cite{PDG2024} &NRQM\cite{Ghalenovi:2014}&NRQM\cite{Shah:EPJA2016}&Error$ (\%) $\\  \hline
&$ \Sigma_{c} $&2.453 &2.455&2.459&2.452&0.08\\
$S$-wave&$ \Xi_{c} $&2.463 &2.465&2.504&2.473&0.08\\
&$\Omega_{c} $&2.694 &2.695&2.566&2.695&0.03\\ \hline

&$ \Sigma_{c} $ &2.791 &2.792&&2.794&0.03\\
$P$-wave&$ \Xi_{c} $& 2.789&2.791&&2.726&0.07\\
&$ \Omega_{c} $ &3.063 &3.065&&2.996&0.06\\  \hline

&$ \Sigma_{c} $& 3.088 && &3.058&\\
$D$-wave&$ \Xi_{c} $ &3.113 &&&2.960&\\ 
&$ \Omega_{c} $&3.325 &&&3.230&\\
\hline

\end{tabular}}
 \end{center}
\end{table}

\begin{table} [h]
\caption{Masses of the ground and excited states of single bottom baryons (in GeV). Last column shows the percentage of relative errors between our results and experimental data.}
\label{tab:mass2}
 \begin{center}
{\begin{tabular}{llllllll} \hline \hline  
State& Baryon&Our result& Exp \cite{PDG2024} &NRQM\cite{Ghalenovi:2014} &NRQM\cite{Thakkar:2017}&RPH\cite{Wei:2017}& Error(\%)\\  \hline
&$ \Sigma_{b} $ & 5.808   & 5.810  & 5.808 & 5.816&5813&0.05\\
$S$-wave&$ \Xi_{b} $ & 5.825&    5.797&5.848  &5.793& 5.793&0.48\\
&$ \Omega_{b} $  &6.043& 6.045 & 5.903  & 6.048& 6.048&0.03\\ \hline

&$ \Sigma_{b} $&6.096 &    6.098 &&6.105& 6.098&0.03\\
$P$-wave&$ \Xi_{b} $&6.213 &    6.227& & 6.133& 6.080&0.22\\
&$ \Omega_{b} $  &6.337&    6.339& &6.321 & 6.325&0.03\\ \hline

&$ \Sigma_{b} $& 6.384&   &   &6.363&6.369& \\ 
$D$-wave&$ \Xi_{b} $&6.352&   &  & 6.355& 6.354&\\ 
&$ \Omega_{b} $ &6.631&   &   &6.554& 6.590&\\
\hline

\end{tabular}}
 \end{center}
\end{table}

We calculated the masses of the ground and $P$-wave and $D$-wave excited states of single and doubly heavy baryons and list them in Tables~\ref{tab:mass1},~\ref{tab:mass2},~\ref{tab:mass3} and~\ref{tab:mass4}. We compare the obtained masses with the exist experimental data~\cite{PDG2024} and those of other theoretical models; non-relativistic quark model (NRQM), relativistic quark model (RQM), heavy diquark effective theory (EFT), heavy quark spin symmetry (HQSS) and Regge phenomenology (RPH)
\cite{Ghalenovi:2014,Ghalenovi:2022uok,Shah:EPJA2016,Thakkar:2017,Wei:2017,Albertus2010,Roberts2008,Yoshida2015,Shah2016,Soto2021,Oritz2023}. In our previous studies within a quark model we solved the Schrödinger equation by a numerical method and consider the heavy baryon states~\cite{Ghalenovi:2014,Ghalenovi:2022uok}. 
Our mass evaluations of the single charm and bottom baryons are listed in Tables ~\ref{tab:mass1} and \ref{tab:mass2}. Our results are in good agreement with the experimental masses. Our predictions for the ground and orbitally excited states of doubly heavy baryons are also summarized in Tables~\ref{tab:mass3} and~\ref{tab:mass4}. We could obtain the experimental mass of the $ \Xi_{cc}^{++}$ doubly heavy baron~\cite{LHCb2018} for the $qcc$ baryon state in Table~\ref{tab:mass3}.

\begin{table} [h]
\caption{Masses of ground states of double charm and bottom baryons (in GeV). }
\label{tab:mass3}
 \begin{center}
{\begin{tabular}{lllllll} \hline \hline  
Baryon& Content&Our results &NRQM\cite{Ghalenovi:2022uok}& NRQM\cite{Albertus2010} &RQM\cite{Roberts2008}&NRQM\cite{Yoshida2015}\\  \hline
$ \Xi_{cc} $&$ qcc $& 3.620&3.679&3.613& 3.676&3.685\\
$ \Xi_{bc} $&$ qbc $& 7.050 &7.003&6.928& 7.020& \\
$ \Xi_{bb} $  &$ qbb $& 10.200&10.325&10.198& 10.340 &10.314\\
$ \Omega_{cc} $&$ scc $&  3.750&3.830 &3.712&3.815&3.832 \\
$ \Omega_{bc} $&$ sbc $&6.900  &7.149  &7.013& 7.147&\\
$ \Omega_{bb} $  &$ sbb $& 10.400&10.466 &10.269&10.456&10.447\\
\hline

\end{tabular}}
 \end{center}
\end{table}

\begin{table}[htbp]
\caption{Masses of excited states of double charm and bottom baryons (in GeV).}
\label{tab:mass4}
 \centering
 \begin{center}
{\begin{tabular}{cccccc} \hline \hline
State&Baryon& Our results &NRQM\cite{Shah2016}&EFT\cite{Soto2021}&NRQM\cite{Oritz2023}\\
\hline
&$ \Xi_{cc} $  & 3.928 &  3.853&4.028&3.855\\  
 &$ \Xi_{bc} $  & 7.266 &  7.140& &\\
 &$ \Xi_{bb} $  & 10.631 & 10.502& 10.386& 10.417 \\
$P$-wave&$ \Omega_{cc} $  &4.106&3.964 & 4.086&4.002 \\
&$ \Omega_{bc} $  & 7.4457& 7.375& &\\
&$ \Omega_{bb} $  &10.808&10.634 &10.607& 10.560\\
\hline
&$ \Xi_{cc} $  & 4.237 & 4.026  &  4.321&  \\
&$ \Xi_{bc} $  & 7.575 & 7.307& & \\
&$ \Xi_{bb} $  & 10.939 & 10.658& 10.585& \\
$D$-wave&$ \Omega_{cc} $  & 4.414& 4.133&4.263& \\
 &$ \Omega_{bc} $  & 7.754&7.807& &\\
&$ \Omega_{bb} $  &11.1177&10.783 &10.723& \\

\hline
\end{tabular}}
 \end{center}
\end{table}

\section{Semileptonic decay widhts of $ B\rightarrow B^{\prime}l\bar{\nu} $ transitions and branching fractions} \label{semi decay}
The semileptonic decays caused by the weak force in which a new hadron, a lepton and a neutrino are produced. To consider the semileptonic decays one needs to determine the transition form factors. There are different methods to simplify the transition form
factors~\cite{Faessler2009,Georgi1990, Carone1991,Flynn2008}. The methodology for the decay widths are similar to the pattern employed in our previous study ~\cite{Ghalenovi:2022dok}.\\
The elements of the semileptonic transition matrix are defined by the weak hadronic and leptonic currents ($ J_{\mu} $ and $ \mathcal{L}^{\mu}$)
\begin{equation}
T=\dfrac{G_F}{\sqrt{2}}V_{cb}J_{\mu}\mathcal{L}^{\mu}
\end{equation}

where $G_F $ and $ V_{bc} $  refers to the Fermi Coupling constant and CKM matrix element. 
\begin{equation}
J_{\mu} = V_{\mu}-A_{\mu}= \bar{\psi}^c\gamma_{\mu}(I-\gamma_5)\psi^b,     \qquad\mathcal{L}^{\mu}=\bar{l}\gamma^{\mu}(1-\gamma_5)\nu_l
\end{equation}

where $ \psi^{b(c)} $ is the charm or bottom quark field and $ V_{\mu}\equiv\bar{\psi}^c\gamma_{\mu}\psi^b $ and $ A_{\mu}\equiv\bar{\psi}^c\gamma_{\mu}\gamma_5\psi^b $. The hadron matrix elements is given as fallows
\begin{equation}
\begin{aligned}
H_{\mu}&= 
\left\langle B^{\prime}(p^{\prime},s^{\prime})\mid V_{\mu}-A_{\mu}\mid B(p,s)\right\rangle=\left\langle B^{\prime}(p^{\prime},s^{\prime})\mid \bar{\psi}^c\gamma_{\mu}(I-\gamma_5)\psi^b\mid B(p,s)\right\rangle
\\
=&\bar{u}^ {\prime}(p^{\prime},s^{\prime})\left\lbrace \gamma_{\mu}(F_1(\omega)-\gamma_5G_1(\omega))+v_{\mu}(F_2(\omega)-\gamma_5G_2(\omega)) \right. \\
& \left. +v^{\prime}_{\mu}(F_3(\omega)-\gamma_5G_3(\omega))\right\rbrace u(p,s)
\end{aligned}
\end{equation}

where $ \mid B(p,s)> $ and $ \mid B^{\prime}(p^{\prime},s^{\prime})> $ refers to the initial and final baryons. $ u(p,s) $ and $ u^{\prime}(p^{\prime},s^{\prime}) $ are Dirac spinors, and $ v_{\mu}=p_{\mu}/m_B (v^{\prime}_{\mu}=p^{\prime}_{\mu}/m_B^{\prime})$ is the four velocity of the corresponding baryon. $F_i(\omega)$ and $G_i(\omega)$ refer to the form factors.
The differential decay rates are given by
\begin{equation}
\dfrac{d\Gamma}{d\omega}=\dfrac{d\Gamma_L}{d\omega}+\dfrac{d\Gamma_T}{d\omega}.
\end{equation}

in which $ \Gamma_L $ and $ \Gamma_T $ are the longitudinally and transversely polarized W's respectively  and obtained in terms of the transition form factors  \cite{Albertus2005,Albertus2008} 

\begin{equation}\label{transverss}
\dfrac{d\Gamma_T}{d\omega}=\frac{G_F^2\vert V_{cb}\vert ^2M_{B^{\prime}}^3}{12\pi^3}\sqrt{\omega^2-1}q^2\lbrace(\omega-1)\vert F_1(\omega)\vert^2+(\omega+1)\vert G_1(\omega)\vert^2\rbrace,
\end{equation}

and

\begin{equation} \label{longitudd}
\dfrac{d\Gamma_L}{d\omega}=\frac{G_F^2\vert V_{cb}\vert ^2M_{B^{\prime}}^3}{24\pi^3}\sqrt{\omega^2-1}\lbrace(\omega-1)\vert \mathcal{F}^V(\omega)\vert^2+(\omega+1)\vert \mathcal{F}^A(\omega)\vert^2\rbrace,
\end{equation}

where

\begin{equation}
\begin{aligned}
\mathcal{F}^{V,A}(\omega)=&\left[ (m_B\pm m_{B^{\prime}})F_1^{V,A}(\omega)+(1\pm\omega) \left( m_{B^{\prime}}F_2^{V,A}(\omega)+m_BF_3^{V,A}(\omega)\right) \right] ,\\&
F_j^V\equiv F_j(\omega), \quad F^A_j\equiv G_j(\omega), \quad j=1,2,3
\end{aligned}
\end{equation}

where the lepton masses are neglected. The velocity transfer $\omega(=v.v^{\prime}) $ takes the values between  $ \omega=1 $ to the maximum of $ \omega_{max}=(m_B^2+m^2_{B^{\prime}})/(2m_B m_{B^{\prime}})$. 
The decay width is obtained as follows

\begin{equation} \label{total gamma}
  \Gamma=\int^{\omega_{max}}_1 d\omega\dfrac{d\Gamma}{d\omega}=\int^{\omega_{max}}_1 d\omega(\dfrac{d\Gamma_L}{d\omega}+\dfrac{d\Gamma_T}{d\omega}).
 \end{equation} 

Different studies have simplified the transition form factors of the semileptonic decays \cite{Faessler2009,Georgi1990, Carone1991,Flynn2008}. Ref.~\cite{Faessler2009} has simplified the form factors employing a Lorentz covariant quark model. In the heavy quark limit and near to zero recoil point, the weak form factors can be simplified and expressed by a single IW function $\eta(\omega)$
 ~\cite{Faessler2009,Isgur1991,Ebert20066} 
 
\begin{eqnarray} \label{formfactors}
F_1(\omega)=G_1(\omega)=\eta(\omega), \quad \quad \quad \quad \quad\quad\\
F_2(\omega)=F_3(\omega)=G_2(\omega)=G_3(\omega)=0 \nonumber.
\end{eqnarray}

\begin{figure}[htbp]
\centering
\includegraphics[width=0.4\textwidth]{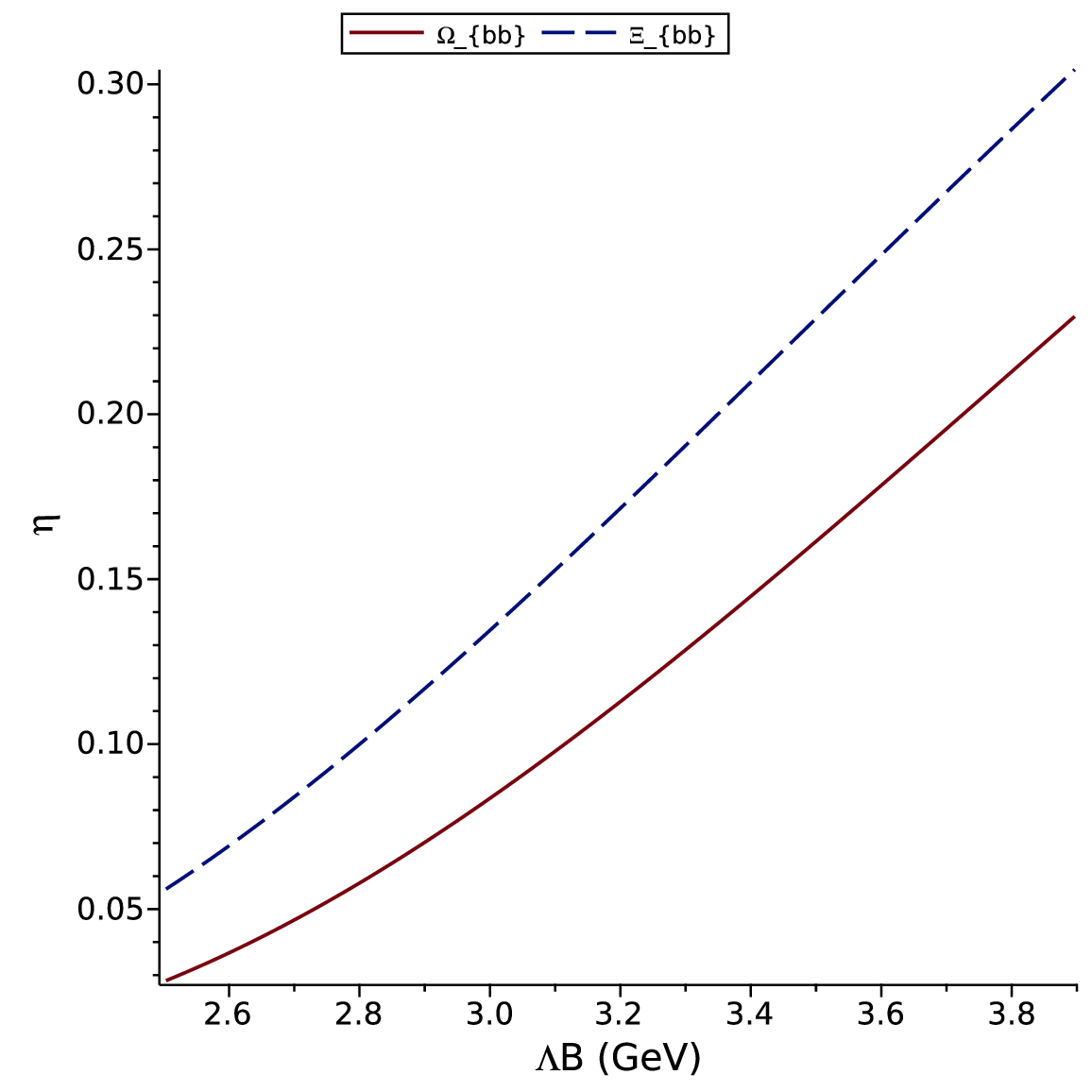}
\caption{$ \eta $ function versus $ \Lambda_B $ for $\Omega_{bb}\rightarrow \Omega_{bc}\ell \bar {\nu}_{\ell}$ and $\Xi_{bb}\rightarrow \Xi_{bc}\ell \bar {\nu}_{\ell}$ transitions ($\ell=e$ or $\mu$). }
\label{fig:1}
\end{figure}

 The universal function $\eta(\omega)$ is defined as  \cite{Faessler2009} 
\begin{equation} \label{eta}
\eta(\omega)=exp\left( -3(\omega-1)\frac{m_{bb}^2}{\Lambda_B^2}\right) 
\end{equation}  

where $m_{bb}=2m_b$ for the $ bb\rightarrow bc $ transitions.  $ \Lambda_B $ depends on the size of a baryon and  ranges from $ 2.5\leqslant \Lambda_B\leqslant 3.5$ GeV \cite{Faessler2009} where, the smaller of $ \Lambda_B $ gives smaller decay widths. In the case of the $ bc\rightarrow cc $ transitions, $ m_{bb}$ should be replaced by $m_{cc} $ and therefore, the IW functions for $ bb\rightarrow bc $ and $ bc\rightarrow cc $ transitions depends on the heavy flavor factors. The $ \Lambda_B $ dependence of $ \eta $ functions for
 $\Omega_{bb} \rightarrow \Omega_{bc}$ and $\Xi_{bb}\rightarrow \Xi_{bc}$
transitions  (with $ \Lambda_B=2.5 $ GeV and $ \omega=\omega_{\rm max} $) is plotted in Fig.~\ref{fig:1}. \\

\begin{figure}[htbp]
\centering
\includegraphics[width=0.4\textwidth]{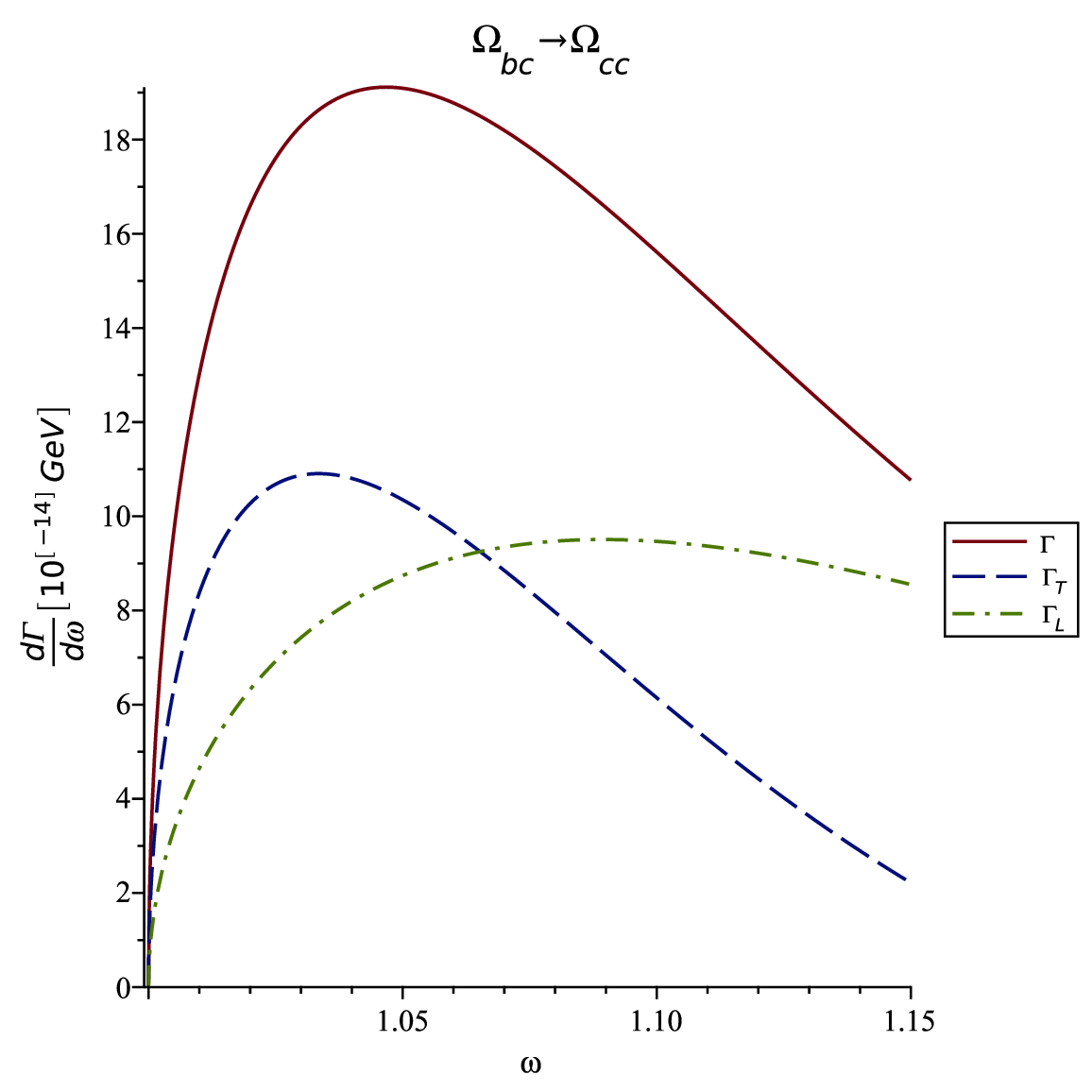}

\caption{$ \frac{d\Gamma_T}{d\omega} $, $  \frac{d\Gamma_L}{d\omega} $ and $ \frac{d\Gamma}{d\omega} $ semileptonic decay widths for  $\Xi_{bc}\rightarrow \Xi_{cc}\ell \bar {\nu}_{\ell}$ transition ($\ell=e$ or $\mu$). }
\label{fig:2}
\end{figure}

\begin{figure}[htbp]
\centering
\includegraphics[width=0.4\textwidth]{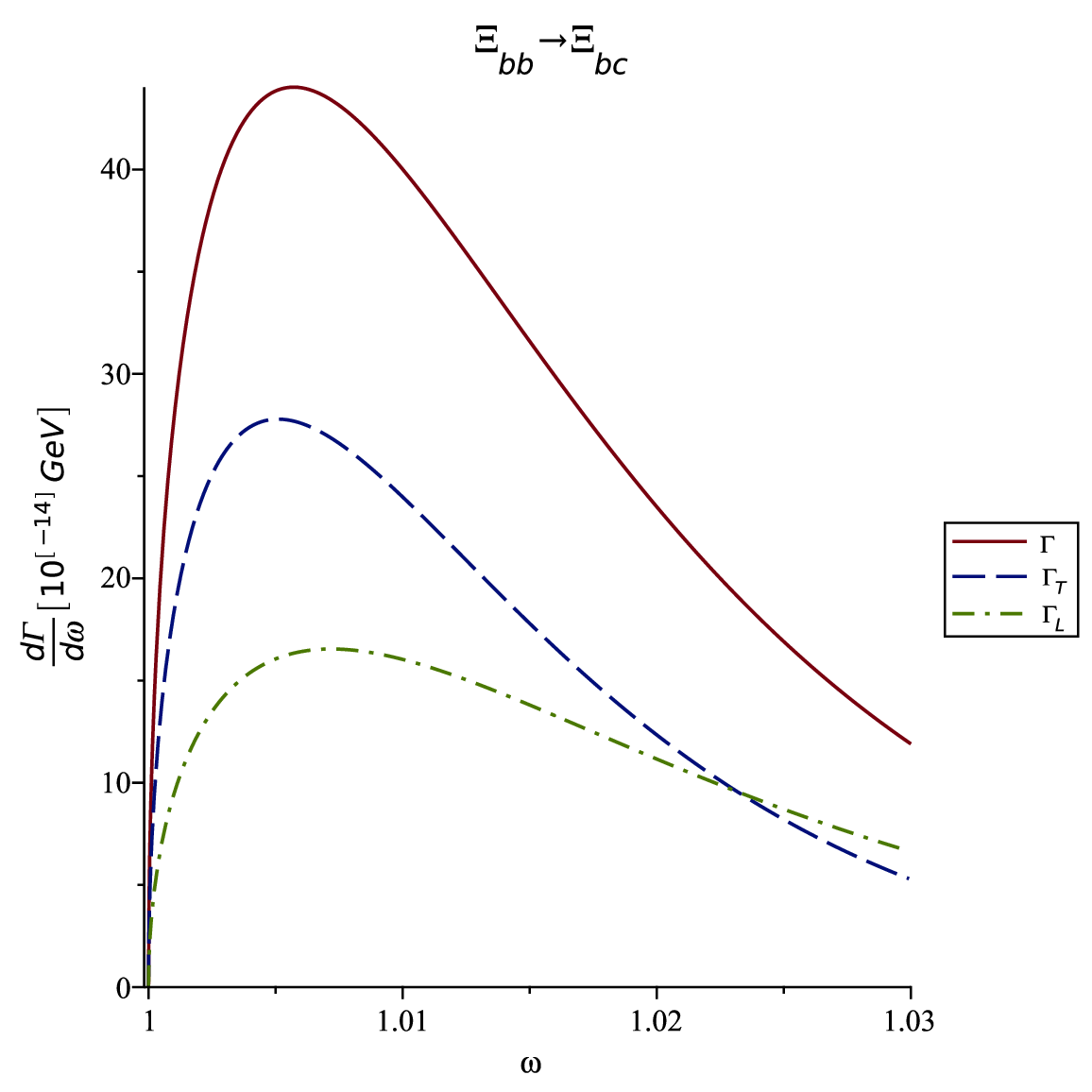}
\caption{ $ \frac{d\Gamma_T}{d\omega} $, $  \frac{d\Gamma_L}{d\omega} $ and $ \frac{d\Gamma}{d\omega} $ semileptonic decay widths for  $\Xi_{bb}\rightarrow \Xi_{bc}\ell \bar {\nu}_{\ell}$ transition  ($\ell=e$ or $\mu$).. }
\label{fig:3}
\end{figure}

The  $ \omega $ dependence of the semileptonic decay rates of $\Xi_{bb}\rightarrow \Xi_{bc}\ell \bar {\nu}_{\ell}$, and $\Xi_{bc}\rightarrow \Xi_{cc}\ell \bar {\nu}_{\ell}$  transitions is shown in Figs.~\ref{fig:2} and ~\ref{fig:3} respectively.

Regarding the equation \ref{formfactors} one can get the following relations
 
\begin{equation}\label{transvers2}
\dfrac{d\Gamma_T}{d\omega}=
\frac{G_F^2\vert V_{cb}\vert ^2M_{B^{\prime}}^3}{6\pi^3}q^2\omega\sqrt{\omega^2-1} \eta^2(\omega),
\end{equation}

\begin{table} [htbp]
    \caption{ Semileptonic decay widths of double charm and bottom  baryons (in units of $10^{-14}$ GeV).}
      \label{tab:semileptonic1}
    \centering
    \begin{center}
    \begin{tabular}{ccccccc}\hline \hline
     Decay~~& ~Our  results~~&NRQM\cite{Ghalenovi:2022dok}&~RCQM~\cite{Faessler2009}~&~NRQM~\cite{hassanabadi2020}~ & ~HQSS~\cite{Hernandez2008}~& ~~LFQM\footnotemark[1]~\cite{Wang2017}~  \\ \hline
$\Xi_{bc}\rightarrow \Xi_{cc}\ell \bar {\nu}_{\ell}$ &4.47$ ^{+1.0}_{-1.20} $ & ~~$ 4.39\pm0.83 $~    &  ~$4.01\pm 1.21 $& 5.07 &2.57$ ^{+0.26}_{-0.03} $&4.50\\
$\Xi_{bb}\rightarrow \Xi_{bc}\ell \bar {\nu}_{\ell}$ &1.66$ ^{+0.73}_{-0.64} $   &~~$1.75\pm 0.73 $~   & ~$1.33\pm 0.61 $  & 1.63& 1.92$ ^{+0.25}_{-0.05} $&3.30\\ \hline
$\Omega_{bc}\rightarrow \Omega_{cc}\ell \bar {\nu}_{\ell}$ &3.55$ ^{+0.67}_{-0.84} $   &~~$4.70\pm0.83$~  &  ~$4.12\pm 1.10 $ & 5.39  &2.59$ ^{+0.20} $ &3.94\\
$\Omega_{bb}\rightarrow \Omega_{bc}\ell \bar {\nu}_{\ell}$ &2.01$ ^{+0.99}_{-0.80} $  & ~~$ 1.87\pm 0.76 $ ~  &~$1.92\pm 1.15 $ &2.48  &2.14$ ^{+0.20}_{-0.02} $ &3.69\\
\hline
    \end{tabular}
    \end{center}
    \footnotetext[1]{LFQM denotes the light-front quark model.}
\end{table}

The branching ratios can be obtained as following
\begin{equation}
{\cal B}=\Gamma\times\tau,
\end{equation}
where $ \tau $ is the lifetime of the initial baryon. We take the values of $ \tau_{\Xi_{bb}}=370\times 10^{-15}s$, $  \tau_{\Xi_{bc}}=244 \times 10^{-15}s $~\cite{karliner2014}, $  \tau_{\Omega_{bc}}=220 \times 10^{-15}s $, and $  \tau_{\Omega_{bb}}=800 \times 10^{-15}s $~\cite{Kiselev2002,Kiselev2002b} to evaluate the branching fractions.

\begin{table}[htbp]
    \caption{The calculated branching ratios of semileptonic decays of double charm and bottom baryons.}
      \label{Branching}
    \centering
    \begin{center}
    \begin{tabular}{ccccc} \hline \hline
     Process~~& Our results~~&  NRQM~\cite{Ghalenovi:2022dok}& NRQM~\cite{hassanabadi2020}~~ & LFQM~\cite{Wang2017}~~ \\
\hline
$\Xi_{bc}\rightarrow \Xi_{cc}\ell \bar {\nu}_{\ell}$~&1.65$\times10^{-2}$ & $1.63\times10^{-2}$ &$1.11\times10^{-2}$&$1.67\times10^{-2}$\\
$\Xi_{bb}\rightarrow \Xi_{bc}\ell \bar {\nu}_{\ell}$~&0.61$\times10^{-2}$ & $0.98\times10^{-2}$ &$0.28\times10^{-2}$&$1.86\times10^{-2}$\\ \hline
$\Omega_{bc}\rightarrow \Omega_{cc}\ell \bar {\nu}_{\ell}$~&1.18$\times10^{-2}$ & $1.57\times10^{-2}$ &$1.10\times10^{-2}$&$1.32\times10^{-2}$\\
$\Omega_{bb}\rightarrow \Omega_{bc}\ell \bar {\nu}_{\ell}$~&2.45$\times10^{-2}$ &$2.27\times10^{-2}$&&$4.49\times10^{-2}$\\
\hline
    \end{tabular}
    \end{center}
\end{table}
 
The evaluated semileptonic decay widths of doubly heavy baryons 
 are listed in Table~\ref{tab:semileptonic1}. We take $\Lambda_{B}=3 $.
 The uncertainties in our predicted widths are due to the
 parameter $\Lambda_{B} $, which varies in the range of $ 2.5-3.5$ GeV. We compare our results with those evaluated by other theoretical works~\cite{Hernandez2008,Wang2017,Ghalenovi:2022dok,Faessler2009,hassanabadi2020}. Our predicted decay widths are close to the ones obtained by Ref.~\cite{Faessler2009}. For the $ \Xi_{bc}\rightarrow \Xi_{cc} $, $\Xi_{bb}\rightarrow
\Xi_{bc} $ and $\Omega_{bb}\rightarrow \Omega_{bc}$ transitions, our results are also in good agreement with those of our previous work~\cite{Ghalenovi:2022dok}. The evaluated branching fractions of semileptonic decays are summarized in Table~\ref{Branching} and compared with other evaluations ~\cite{Wang2017,Ghalenovi:2022dok,hassanabadi2020}. We hope our results are helpful to find the value of the CKM matrix element $ V_{cb} $ from next experiments. 
 
\section{Summary} \label{Summary}
In the present work, we propose our phenomenological approach for estimation of energy eigenvalues of three-body baryon systems by integrating the deep learning with particle swarm optimization methods. By integrating deep learning with PSO, we could develop a powerful methodology for accurate energy prediction of three-body quantum systems. The deep learning model, trained on a large dataset generated using the shooting method, provides an initial estimate of the energy values while, PSO refines these predictions and leads to highly accuracy in calculations. This hybrid approach demonstrates the potential of combining machine learning and optimization techniques in solving challenging math problems in computational physics.
 By evaluation of baryon energies, we are able to predict the mass spectrum of the single and doubly heavy baryons which are not yet determined by experiment. The validity of our scheme for heavy baryon systems will be determined by consistency with the exist experimental data. Our evaluated masses are in a good agreement with the experiments and also with the predictions of other theoretical works. In our study the overall MSE, training MSE and validation MSE are $6.443\times 10^{-5}$, $6.690\times 10^{-5}$ and $6.196\times 10^{-5}$ respectively.\\
The computed errors in the last columns of Tables \ref{tab:mass1} and \ref{tab:mass2} shows that our model is successful in describing the baryon systems. For the D-wave excited states of single charm and bottom baryons there is no experimental data to make a comparison between the results. For the  D-wave $ \Sigma_{c} $, $ \Sigma_{b} $, $ \Xi_{c} $ and $ \Xi_{b} $ heavy baryon states our results are very close to the other predictions but, for the $ \Omega_{c} $ and $ \Omega_b $ baryon states our evaluated masses are higher than those of other works. For the P-wave and D-wave doubly heavy baryon states, all of our predictions are higher than the masses of other works especially,  for the D-wave $ \Omega_{bb} $ baryon states our evaluated mass is about 400 MeV higher than the masses predicted in Refs. \cite{Shah2016} and \cite{Soto2021}.  For the ground state doubly heavy baryons, our results are very close to the baryon masses obtained in Ref. \cite{Albertus2010}. The obtained heavy baryon masses and decay widths of the present work are very close to the predictions obtained in our previous works \cite{Ghalenovi:2022uok} and \cite{Ghalenovi:2022dok} in which we used a non-relativistic hypercentral quark model.  Using our obtained masses we can get the mass differences $ \Delta_M$ between the ground state $ \Omega_{cc} $  and the corresponding $ \Xi_{cc} $ doubly heavy baryons. We get  $ \Delta_M=150$  MeV  which is very close to the  $ \Delta_M=155\sim158 $  MeV predicted in Ref.~\cite{Ebert2004}. Limited number of theoretical models could evaluate the both of the single and doubly heavy baryon masses in one study. Using the deep learning model we could obtain a good mass spectrum for the both of the single and doubly charm and bottom baryons in the present study. 
 Finally, We study the semileptonic decays of doubly heavy $\Xi$ and $\Omega$ baryons. We introduce a
 new form of the IW function, and evaluate the semileptonic decay widths and branching
ratios of the $ bb\rightarrow bc $ and $ bc\rightarrow cc $ transitions.
A comparison between our theoretical results and other available predictions shows our obtained results are acceptable. Our predictions will be able to guide the future searches for the undiscovered single and doubly heavy baryons at LHCb, ATLAS, and CMS.

\end{document}